\documentclass[usenatbib]{mn2e}

\usepackage{graphicx}
\usepackage{amssymb}
\usepackage{soul}

\newcommand{\aj}{AJ}
\newcommand{\apj}{ApJ}
\newcommand{\apjl}{ApJ}
\newcommand{\apjs}{ApJS}
\newcommand{\aap}{A\&A}
\newcommand{\mnras}{MNRAS}
\newcommand{\nat}{Nature}
\newcommand{\gcn}{GCN Circular}
\newcommand{\atel}{ATel}

\newcommand{\pasj}{PASJ}
\newcommand{\apss}{Ap\&SS}

\newcommand{\na}{New A}
\newcommand{\araa}{ARA\&A}%

\newcommand{\arcdeg}{\ensuremath{^{\circ}}}
\newcommand{\Msun}{\ensuremath{M_{\odot}}}
\newcommand{\nH}{\ensuremath{N_{{\rm H}}}}

\newcommand{\swiftseventeen}{Swift\,J1745$-$26}
\newcommand{\swiftseventeenlong}{Swift\,J174510.8$-$262411}

\newcommand{\swift}{\emph{Swift}}
\newcommand{\integral}{\emph{INTEGRAL}}

\newcommand{\lmxb}{LMXB}
\newcommand{\lmxbs}{LMXBs}

\title
[The evolving polarised jet of \swiftseventeen]
{The evolving polarised jet of black hole candidate \swiftseventeen}

\author[P.A.~Curran et al.]
{P.A.~Curran$^{1}$\thanks{e-mail: peter.curran@curtin.edu.au},
M.~Coriat$^{2}$,
J.C.A.~Miller-Jones$^{1}$, 
R.P.~Armstrong$^{2,3}$,\newauthor
P.G.~Edwards$^{4}$,
G.R.~Sivakoff$^{5,6}$,
P.~Woudt$^{2}$ 
D.~Altamirano$^{7}$,
T.M.~Belloni$^{8}$,\newauthor
S.~Corbel$^{9,10}$,
R.P.~Fender$^{11}$,
E.G.~K\"{o}rding$^{12}$,
H.A.~Krimm$^{13,14}$,
S.~Markoff$^{7}$,\newauthor
S.~Migliari$^{15}$,
D.M.~Russell$^{16,17}$,
J.~Stevens$^{18}$ and
and T.~Tzioumis$^{4}$\\
$^{1}$International Centre for Radio Astronomy Research, Curtin University, GPO Box U1987, Perth, WA 6845, Australia\\ 
$^{2}$Department of Astronomy, University of Cape Town, Private Bag X3, Rondebosch 7701, South Africa\\ 
$^{3}$SKA South Africa, 3rd Floor, The Park, Park Road, Pinelands, 7405, South Africa \\ 
$^{4}$CSIRO Astronomy and Space Science, Australia Telescope National Facility, P.O. Box 76, Epping, NSW 1710, Australia \\ 
$^{5}$Department of Physics, University of Alberta, CCIS 4-183 Edmonton, AB T6G 2E1, Canada \\ 
$^{6}$Department of Astronomy, University of Virginia, P.O. Box 400325, Charlottesville, VA 22904, USA\\ 
$^{7}$Astronomical Institute Anton Pannekoek, University of Amsterdam, Science Park 904, 1098XH Amsterdam, The Netherlands \\ 
$^{8}$INAF -- Osservatorio Astronomico di Brera, Via E. Bianchi 46, I-23807, Merate, Italy \\ 
$^{9}$Laboratoire AIM (CEA/IRFU--CNRS/INSU--Universit\'e Paris Diderot), CEA DSM/IRFU/SAp, F-91191 Gif-sur-Yvette, France\\ 
$^{10}$Institut Universitaire de France, F-75005 Paris, France \\ 
$^{11}$School of Physics and Astronomy, University of Southampton, Southampton, Hampshire, SO17\,1BJ, UK \\ 
$^{12}$Radboud Universiteit Nijmegen, IMAPP, 6525KL Nijmegen, The Netherlands \\ 
$^{13}$Universities Space Research Association, Columbia, MD 21044, USA \\ 
$^{14}$NASA/Goddard Space Flight Center, Astrophysics Science Division, Code 661, Greenbelt, MD 20771, USA  \\ 
$^{15}$Departament d'Astronomia i Meteorologia, Universitat de Barcelona, Mart\'i~i Franqu\`es 1, 08028 Barcelona, Spain \\ 
$^{16}$Instituto de Astrof\'isica de Canarias (IAC), E-38200 La Laguna, Tenerife, Spain \\ 
$^{17}$Departamento  de Astrof\'isica, Universidad de La Laguna (ULL), E-38206 La Laguna, Tenerife, Spain  \\ 
$^{18}$CSIRO Astronomy and Space Science, Australia Telescope National Facility, Locked Bag 194, Narrabri NSW 2390, Australia \\ 
}

\begin{document}

\date{Accepted 2013 October 31. Received 2013 September 16}

\pagerange{\pageref{firstpage}--\pageref{lastpage}} \pubyear{}

\maketitle

\label{firstpage}


\begin{abstract}
\swiftseventeen\ is an X-ray binary towards the Galactic Centre that
was detected when it went into outburst in September 2012. This source
is thought to be one of a growing number of sources that display
``failed outbursts'', in which the self-absorbed radio jets of the
transient source are never fully quenched and the thermal emission
from the geometrically-thin inner accretion disk never fully dominates
the X-ray flux.
We present multifrequency data from the Very Large Array, Australia
Telescope Compact Array and Karoo Array Telescope (KAT- 7) radio
arrays, spanning the entire period of the outburst.
Our rich data set exposes radio emission that displays a high level of
large scale variability compared to the X-ray emission and deviations
from the standard radio--X-ray correlation that are indicative of an
unstable jet and confirm the outburst's transition from the canonical
hard state to an intermediate state.
We also observe steepening of the spectral index and an increase of
the linear polarization to a large fraction ($\approx 50$\%) of the total flux, as well
as a rotation of the electric vector position angle. These are
consistent with a transformation from a self-absorbed compact jet to
optically-thin ejecta -- the first time such a discrete ejection has
been observed in a failed outburst -- and may imply a complex magnetic
field geometry.
\end{abstract}



\section{Introduction}\label{section:introduction}

Once thought to be an anomaly, relativistic jets are now accepted to
be a standard feature of stellar mass black holes in actively
accreting low mass X-ray binary (\lmxb) systems (e.g.,
\citealt{Mirabel1998:Natur.392,Fender2006:csxs.book381}), possibly
remaining active in quiescence (e.g., \citealt{Gallo2006:MNRAS.370}).
During outburst, powered by increased accretion onto the black hole,
the radio jets evolve through various phases: initially increasing in
power before being quenched and, later, reactivated. The morphology of
the radio jets depends on \citep{Fender2006:csxs.book381} the observed
X-ray ``states'' (see e.g.,
\citealt{mclintock2006:csxs157,belloni_2010LNP...794}). In the {\it
  hard} state, when the X-ray spectrum is dominated by power-law
emission from the optically-thin, geometrically-thick inner regions,
the radio jets are described by self-absorbed synchrotron emission
with a flat ($\alpha \sim 0$) or inverted spectrum ($\alpha > 0$),
where $F_{\nu} \propto \nu^{\alpha}$; this is interpreted as an
optically-thick, compact jet.  In the {\it soft} or {\it
  thermal-dominant} state, when the X-ray spectrum is dominated by a
thermal blackbody component from the accretion disk that extends down
to the innermost stable circular orbit (ISCO), the radio jets are
observed to be quenched by a factor of at least 700 (e.g.,
\citealt{Russell2011:ApJ.739,Coriat2011:MNRAS.414}). In the transition
between hard and soft, defined by various classes of {\it
  intermediate} states, the radio is in some cases observed to become
optically thin ($\alpha < 0$) and to exhibit flares; in a number of
sources, these have been spatially resolved (e.g.,
\citealt{Tingay1995:Natur.374,Mirabel1998:Natur.392,Miller-Jones2012:MNRAS.421})
as discrete ejecta.

The connection between the radio and X-ray properties of \lmxbs\ is
further demonstrated by the strong correlation that exists between
their X-ray and radio luminosities in the hard state (e.g.,
\citealt{Hannikainen1998:A&A.337,Corbel2000:A&A.359,Corbel2003:A&A.400,Gallo2003:MNRAS.344,Corbel2013:MNRAS.428}).
In recent years a number of outliers to this correlation have been
identified -- including XTE\,J1859+226
\citep{Corbel2004:ApJ.617}, IGR\,J17497$-$2821
\citep{Rodriguez2007:ApJ.655}, \swift\,J1753.5$-$0127
\citep{Cadolle-Bel2007:ApJ.659}, H1743$-$322
\citep{Coriat2011:MNRAS.414}, MAXI\,J1659$-$152
\citep{Jonker2012:MNRAS.423} and XTE\,J1752$-$223
\citep{Ratti2012:MNRAS.423} -- which called into question the
universality of the relationship.
However, it is now accepted that these outliers form a distinct
population \citep{Gallo2012:MNRAS.423} that form a secondary (radio
quiet) branch obeying its own correlation. Furthermore, at low
luminosities a transition is observed between the two branches
\citep{Coriat2011:MNRAS.414,Jonker2012:MNRAS.423,Ratti2012:MNRAS.423}.
Interestingly, it has recently been suggested that AGN might show the
same bimodal correlation \citep{King2013ApJ.774}, supporting the idea
that the same physical mechanisms extract energy from black holes at
very different mass scales.

How and when the quenching and reactivation of the jets occurs during
the intermediate states, and what causes the discrete ejecta are still
poorly understood. In fact, in an increasing number of ``failed
outbursts'', the system never transitions to the soft state, the
self-absorbed radio jets are never fully quenched, and the
geometrically-thin accretion disk likely remains truncated at a radius
greater than the ISCO and never fully dominates the observed X-ray
flux (e.g., 9 sources in \citealt{Brocksopp2004:NewA.9} and references
therein;
\citealt{Wijnands2002:ApJ.564,Capitanio2009:MNRAS.398,Ferrigno2012:A&A.537,Soleri2013:MNRAS.429}).
In some outbursts multiple radio flares are observed when the jet is
repeatedly suppressed and reactivated in the intermediate states
before, or instead of, a full state transition
\citep{Fender2004:MNRAS.355}.  Such outbursts may reveal new
information on jet suppression and reactivation in both failed and
successful outbursts of X-ray binaries.

Radio jets are normally described by their photometric, spectral and,
occasionally, spatial properties, as summarised above, but only a few have
been observed to exhibit polarization, which can be used to infer
physical properties. While optically thick and thin synchrotron
emission from the jet can, in the presence of an ordered magnetic
field, produce linear polarizations of up to $\approx 10\%$ or
$\approx 70\%$, respectively \citep{Longair1994:hea2.book}, relatively
few \lmxbs\ have been observed to do so (e.g.,
\citealt{Fender2003:Ap&SS.288} and references therein;
\citealt{Brocksopp2007:MNRAS.378,Brocksopp2013:MNRAS.432}).  If
detected, polarization can be used to infer properties of the jet,
such as orientation, and of the magnetic field of the jet and the
surrounding medium (e.g., \citealt{Stirling2004:MNRAS.354}). However,
the magnetic field is not necessarily ordered and a number of
mechanisms -- such as multiple, unresolved components that cancel each
other out, or spatially dependent Faraday rotation -- can suppress the
resulting net the polarization (see e.g.,
\citealt{Brocksopp2007:MNRAS.378} and references therein).

\subsection{\swiftseventeen}\label{sec:intro:source}

The transient black hole candidate \swiftseventeen\ (also known as
\swiftseventeenlong) was discovered in the Galactic Centre region
($l,b = 2.11\arcdeg, 1.40\arcdeg$) by the Burst Alert Telescope (BAT)
on board the \swift\ satellite on September 16 2012 at 09:16 UT (MJD
56186.38618) \citep{Cummings2012:GCN.13774}.  The narrow-field
instruments on \swift\ -- the X-ray Telescope (XRT) and the
Ultraviolet/Optical Telescope (UVOT) -- began observing the source
approximately 1260 seconds after the trigger and detected an X-ray
counterpart \citep{Cummings2012:GCN.13775,Sbarufatti2012:ATel.4383}.
While no optical/UV source was detected by \swift\ (unsurprising given
that the extinction in that direction of $E_{B-V} \approx 3.3$
\citep{schlegel1998:ApJ500} implies $\gtrsim10$ magnitudes
\citep{cardelli1989:ApJ345} of attenuation in any of the
\swift\ observing bands), an infrared counterpart was identified on
the basis of variability against archival images
\citep{Rau2012:ATel.4380}.  It was suggested that the source was an
\lmxb\ black hole system on the basis of X-ray (\swift\ and \integral)
spectral and timing observations. These were also used to classify its
state at various epochs throughout the outburst as being either hard
or hard-intermediate (e.g.,
\citealt{Belloni2012:ATel.4450,Grebenev2012:ATel.4401,Tomsick2012:ATel.4393,Vovk2012:ATel.4381,Sbarufatti2013:ATel.4782})
suggesting that this was a ``failed outburst''.  The black hole nature
of the source is further supported by optical observations during
outburst which display a broadened H$_{\alpha}$ emission line
indicative of a black hole accretor
\citep{Munoz-Darias2013:MNRAS.432}.

Observations at the Karl G. Jansky Very Large Array (VLA) localised
the position of the detected radio counterpart to be RA = 17:45:10.849
$\pm$ 0.001, Dec = $-$26:24:12.60 $\pm$ 0.01\footnote{All
  uncertainties in this paper are quoted and/or plotted at the
  $1\sigma$ confidence level.} (J2000;
\citealt{Miller-Jones2012:ATel.4394}) while observations at the
Australia Telescope Compact Array (ATCA) found a spectral index
consistent with optically-thick synchrotron emission from a partially
self-absorbed compact jet \citep{Corbel2012:ATel.4410}.
In this article we present the full sets of radio data from ATCA, VLA
and Karoo Array Telescope (KAT-7) monitoring observations of
\swiftseventeen\ obtained over the period of the outburst.
In section \ref{section:observations} we introduce the observations
and reduction methods, while in section \ref{section:discussion} we
discuss the results of our photometric and spectral analyses of the
data and discuss their physical implications for the system. We
summarise our findings in section \ref{section:conclusion}.


\section{Observations \& Analysis}\label{section:observations}

\subsection{Radio data}\label{section:radio}

\begin{figure*} 
  \centering 
  \resizebox{16cm}{!}{\includegraphics[angle=0]{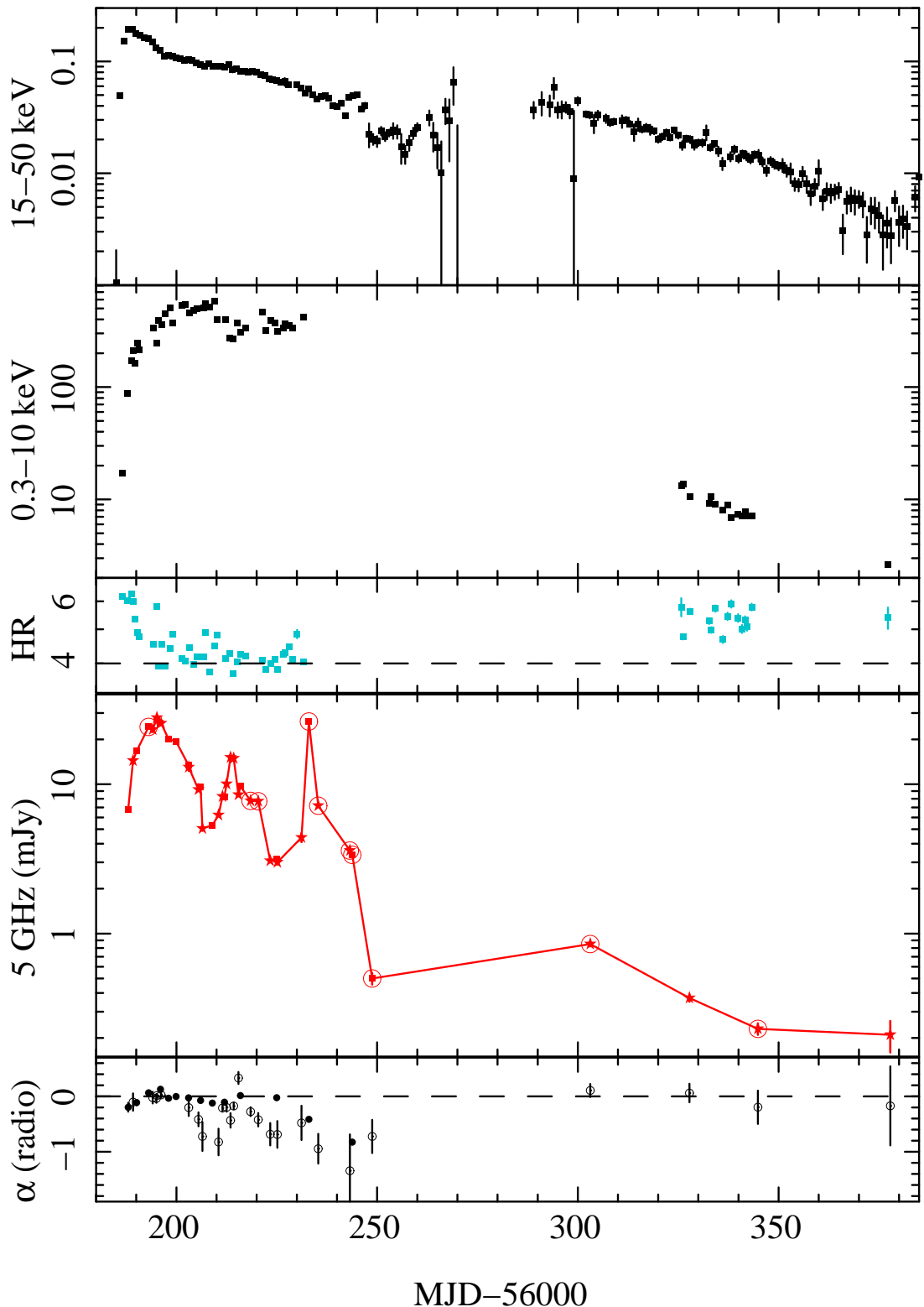} }
  \caption{Hard (BAT; 15--50\,keV) and soft (XRT; 0.3--10\,keV) X-ray
    light curves (counts/second), the soft X-ray hardness ratio
    (1.5--10\,keV/0.3--1.5\,keV), the 5 and 5.5\,GHz radio light
    curves (mJy; squares and stars respectively), and the radio
    spectral index over the period of our observations. The dashed
    horizontal lines on the hardness ratio and radio spectral index
    are purely to guide the eye as to the relative behaviour
    throughout the outburst. Connecting lines are used to demonstrate
    the general behaviour of the radio light curve but we caution that
    it is highly dependent on sampling; the encircled radio fluxes are
    the epochs that were not included on the correlation plot
    (Figure\,\ref{fig:Lx-Lr}) due to a lack of quasi-simultaneous
    X-ray flux. The radio spectral indices represented by hollow
    circles are epochs with only 2 observed frequencies, and whose
    errors may therefore be underestimated.}
 \label{fig:broad_lc} 
\end{figure*}

\begin{figure*} 
  \centering 
  \resizebox{16cm}{!}{\includegraphics[angle=0]{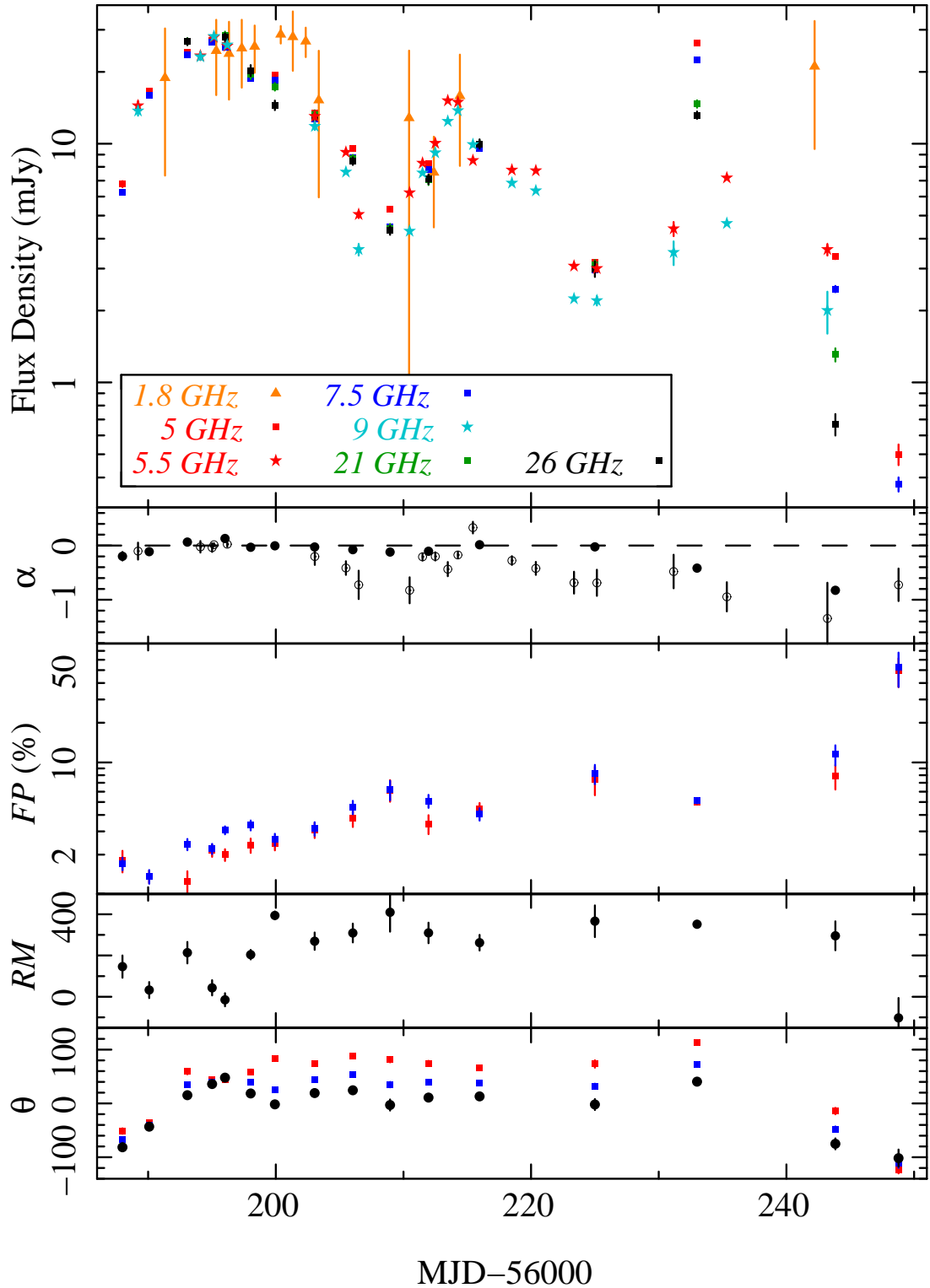} }
  \caption{Radio light curves, spectral, and polarization parameters
    of the source during outburst. The upper panel shows the flux
    densities at each of the 7 radio frequencies observed (legend is
    common to all panels; VLA observations are represented by squares,
    ATCA observations by stars, and KAT-7 by triangles), while the
    second panel shows the radio spectral index (as in
    Figure\,\ref{fig:broad_lc} the spectral indices represented by
    hollow circles are epochs with only 2 observed frequencies). The
    third panel shows the fractional polarizations, {\it FP}, at 2 of
    those frequencies (5GHz and 7.5GHz). The fourth and fifth panels
    show the rotation measure, $RM$, and intrinsic electric vector
    position angle (EVPA; black circles), $\theta$, derived from the
    observed polarization angles (shown in blue and red in the final
    panel).
}
 \label{fig:Radio_curves} 
\end{figure*}

\subsubsection{VLA}\label{section:vla}

\swiftseventeen\ was observed by the VLA from September 18 to November
17 2012 (16 epochs in the most extended, A, configuration) at multiple
frequency bands from 1--48\,GHz (Table\,\ref{table:vla}), though on
the majority of the epochs the source was only observed up to 26\,GHz.
Each frequency was comprised of 2 basebands, with 8 spectral windows
of 64 2\,MHz channels each -- giving a bandwidth of 1.024\,GHz per
baseband. The exception is the 1.5\,GHz band, which has 16 spectral windows
of 64 1\,MHz channels each, again giving a bandwidth of 1.024\,GHz.
Flagging, calibration and imaging of the data followed standard
procedures and was carried out within the Common Astronomy Software
Application (CASA) package \citep{McMullin2007:ASPC376}.  The
1.5\,GHz band is affected by a high level of radio frequency
interference (RFI) which reduces the usable bandwidth to 512\,MHz,
spread over the baseband.

The primary calibrator used as the bandpass and polarization angle
calibrator, and to set the amplitude scale in all bands was 3C286
(a.k.a. J1331+3030).  The choice of secondary calibrator was dependent
on the frequency of the observations, with J1751$-$2524 being used at
frequencies $<10$GHz, J1744$-$3116 being used at frequencies from
$10-20$GHz, and J1745$-$2900 being used at frequencies $>20$GHz.  The
polarization leakage calibrator, where polarization calibration was
performed, was J1407+2827.
Images were made and phase self-calibration was then performed on a
per-baseband basis. Due to the large fractional bandwidth
($\Delta\nu/\nu$) of the 1.5, 5.0, and 7.5 GHz bands, additional self-calibration
on a per-spectral window basis was applied.

The flux densities of the source were measured by fitting a point
source in the image plane (Stokes $I$), and, as is usual for VLA data,
systematic errors of 1\% ($<10$GHz), 3\% ($10-40$GHz), or 5\%
($>40$GHz) were added. In the 1.5, 5.0, 7.5 14.4 and 17.2 GHz bands,
Stokes $Q$ and $U$ fluxes were also measured at the position of peak
flux (Table\,\ref{table:fluxes}).
Due to the birefringence of the local or interstellar medium, Faraday
rotation (see section \ref{section:polarization}) will cause a
rotation of the linear polarization vectors. Over the wide fractional
bandwidth of the 1.5\,GHz band this will cause a smearing out of any
polarization signal over each 64\,MHz spectral window; hence the
Stokes $Q$ and $U$ fluxes must be extracted on a per-channel basis.

\begin{table}	
  \centering	
  \caption{Band names, central frequencies, bandwidths and systematic errors of the radio observations.}
\label{table:vla} 	
\begin{tabular}{l l l l} 
  \hline
  Band & Freq. &  Bandwidth &  Systematic  \\
       & (GHz) &  (MHz)     & (\%) \\
  \hline
  VLA: & & & \\
  L      & 1.5  & 1024 & 1 \\ 
  C$_{5}$ &  5.0 & 1024 & 1  \\ 
  C$_{7}$ &  7.5 & 1024 & 1   \\ 
  U$_{14}$ & 14.4   & 1024 & 3  \\ 
  U$_{17}$ & 17.2   & 1024 & 3   \\  
  K$_{21}$ & 20.8   & 1024 & 3   \\ 
  K$_{26}$ & 25.9   & 1024 & 3  \\ 
  Ka$_{32}$ & 31.5  & 1024 & 3  \\ 
  Ka$_{38}$ & 37.5  & 1024 & 3   \\ 
  Q$_{42}$ & 41.5   & 1024 & 5   \\ 
  Q$_{48}$ & 47.5   & 1024 & 5   \\ 
  KAT-7: & & & \\
        & 1.8  & 234 & ... \\
  ATCA: & &  & \\
       & 5.5  & 2048  & 1 \\
       & 9.0   & 2048 & 1 \\
  \hline 
\end{tabular}
\end{table}

\subsubsection{ATCA}\label{section:atca}

The ATCA carried out a long term monitoring campaign on
\swiftseventeen\ at 5.5 and 9 GHz during 24 epochs from September 19
2012 to March 27 2013. Observations were carried out using the Compact
Array Broadband Backend (CABB, \citealt{Wilson2011:MNRAS.416}) with
the array in a number of configurations ranging from H168 to 6A.  Each
frequency band was composed of 2048 channels of 1 MHz. We used
PKS\,1934$-$638 for absolute flux and bandpass calibration, and
J1710$-$269 to calibrate the antenna gains as a function of
time. Flagging, calibration and imaging were carried out with the
Multichannel Image Reconstruction, Image Analysis and Display (MIRIAD)
software \citep{Sault1995:ASPC77}. The flux densities of the source
were measured by fitting a point source in the image plane
(Table\,\ref{table:fluxes}).  Systematic errors for the ATCA fluxes
are approximately 1\%.

\begin{table*}
  \centering	
  \caption{Sample radio flux densities of source, $F_{\nu}$, at
    frequency, $\nu$, and Stokes $Q$ and $U$ flux densities at that
    frequency (all given before systematic errors are added). Full,
    plain-text table is available online.}
\label{table:fluxes} 
\begin{tabular}{l l l l l} 
  \hline
  Epoch   & $\nu$  & $F_{\nu}$     & $Q$     & $U$  \\
  (MJD)   & (GHz) & (mJy) & (mJy beam$^{-1}$) & (mJy beam$^{-1}$)  \\
  \hline
56187.991 & 5.0 & 6.77 $\pm$ 0.13   & -0.025 $\pm$ 0.021 & -0.113 $\pm$ 0.021 \\ 
56187.991 & 7.5 & 6.26 $\pm$ 0.03   & -0.075 $\pm$ 0.012 & -0.074 $\pm$ 0.012 \\ 
56189.210 & 5.5 & 14.40 $\pm$ 0.70   & ... & ...  \\ 
56189.210 & 9.0 & 13.70 $\pm$ 0.60   & ... & ...  \\ 
56190.087 & 5.0 & 16.66 $\pm$ 0.06   & 0.063 $\pm$ 0.022 & -0.206 $\pm$ 0.024 \\ 
56190.087 & 7.5 & 15.92 $\pm$ 0.07   & 0.035 $\pm$ 0.023 & -0.205 $\pm$ 0.025 \\ 
56191.327 & 1.90 & 19 $\pm$ 12   & ... & ...  \\ 
\hline
\end{tabular}
\end{table*}

\subsubsection{KAT-7}\label{section:kat7}

Observations with the 7-dish MeerKAT test array, KAT-7 (for further
details see \citealt{Armstrong2013:MNRAS.433}) were performed at a
central frequency of 1.822\,GHz during 13 epochs from September 13 to
November 11 2012. The maximum and minimum baselines of the array are
192\,m and 24\,m respectively. PKS\,1934$-$638 was used as the primary
(bandpass and flux) calibrator while J1713$-$2658 was used as the
secondary (gain and phase) calibrator during all epochs. The 234\,MHz
bandwidth was made up of 600 channels, each 390.625\,kHz wide.

Visibility data were flagged with AOFlagger\footnote{{\tt
    http://sourceforge.net/projects/aoflagger/}}, which removed an
average of 1.9\% of the recorded data due to Radio Frequency
Interference (RFI), before calibration and image analysis were
performed with the CASA radio astronomy package
\citep{McMullin2007:ASPC376}.  Source flux densities
(Table\,\ref{table:fluxes}) were obtained by subtracting the quiescent
field (observed on June 6 2013, after the source had faded at radio
frequencies) in the image plane, then measuring the residual at each
epoch, as detailed in \cite{Armstrong2013:MNRAS.433}.

\subsubsection{OVRO}\label{section:ovro}

We attempted to observe the source at 15\,GHz with the Owens Valley
Radio Observatory (OVRO) 40\,m telescope, within the regular OVRO
blazar monitoring program \citep{Richards2011:ApJS.194}.  However, all
observations were non-detections or unreliable due to the low
declination and low observing elevation of the source (T. Hovatta,
personal communication).

\subsection{X-ray data}\label{section:xray}

For the purpose of comparison, \swift\ X-ray light curves were
obtained from the XRT online tool\footnote{{\tt
    http://www.swift.ac.uk/user\_objects/}}
\citep{Evans2009:MNRAS.397} and the BAT transient
monitor\footnote{{\tt
    http://swift.gsfc.nasa.gov/docs/swift/results/transients/}}
\citep{Krimm2013:ApJS.209}. Both these resources offer count rate
light curves (and, for the XRT, hardness ratios defined by the ratio
of 1.5--10 keV/0.3--1.5 keV count rates) extracted using standard
procedures (Figure\,\ref{fig:broad_lc}). The un-sampled period of
X-ray data is when the source was unobservable due to the position of
the Sun. Due to the source's proximity to the Galactic Centre, the
MAXI X-ray monitor aboard the International Space Station
\citep{Matsuoka2009:PASJ.61} was unable to obtain data on the source
during the outburst.

To calculate X-ray luminosities, via fluxes, in the 1--10\,keV and
3--9\,keV ranges, average XRT spectra were obtained from the XRT
online tool. The periods of the spectra were chosen so that they
coincided with the initial, rising period of the XRT light curve (MJD
56180--56200), the softest period of the outburst (MJD 56200--56250)
and the late-time period when the hardness ratio had recovered to its
original value (MJD 56320--56380). Only at early times -- when the
hardness ratio, and hence spectra, were obviously evolving -- will
using an average spectrum cause an error, but any such error will be
negligible compared to other uncertainties in estimating the
luminosities. The spectra were fit with a simple absorbed power law in
{\tt XSPEC} (photon index, $\Gamma \approx 1.5\mbox{--}2.5$, column
density, $\nH \approx 1\mbox{--}1.4 \times 10^{22}$, in agreement with
the automated fits of the online tool); including a black body
component in the fit made no discernible improvement. Count rate to
unabsorbed flux conversions of $2.1\times 10^{-10}$, $3.4\times
10^{-10}$ and $1.1\times 10^{-10}$ erg\,cm$^{-2}$\,count$^{-1}$
(1--10\,keV) or $1.1\times 10^{-10}$, $1.2\times 10^{-10}$ and
$0.6\times 10^{-10}$ erg\,cm$^{-2}$\,count$^{-1}$ (3--9\,keV) were
derived from the average spectra over the three epochs.  A more
complete treatment of the X-ray data is beyond the scope of this work
and is the subject of another work (Sbarufatti et al., in
preparation).


\subsection{Spectral indices}\label{section:indices}

We fit (linearly in log-log space) the derived radio flux densities,
$F_{\nu}$, against frequency, $\nu$, to obtain the spectral index,
$\alpha$, of the radio spectra at each epoch. The flux density of each
band was used, except in the case of the 1.5\,GHz band where flux
densities per 64\,MHz spectral window were used, but we obtain similar
values if we also use the per-spectral window flux densities at 5.0
and 7.5 GHz. All epochs were reasonably well fit by a single power law
with no need for additional components.
On many of the epochs, only 2 radio bands were observed (either 5 and
7.5 GHz or 5.5 and 9 GHz) and on these days the spectral index is
under-constrained by the data and hence, may be less accurate.  In the
bottom panel of Figure\,\ref{fig:broad_lc} and the second panel of
Figure\,\ref{fig:Radio_curves} we have therefore drawn a distinction
between these points and those that used a broader fit to measure the
index.

\subsection{Polarization}\label{section:polarization}

The polarization parameters were derived from the measured flux
densities of the Stokes $Q$ and $U$ images: 
linear polarization, $LP = \sqrt{Q^2 + U^2}$;
fractional polarization, $FP = 100 \sqrt{Q^2 + U^2}/I$;
and polarization angle, $PA = 0.5 \arctan(U/Q)$, which is degenerate such
that derived angles may be offset by an integer  multiple of $\pm
180$\arcdeg\ from the true value.
The derived polarimetric parameters at 5.0, 7.5, 14.4 and 17.2 GHz are
presented in Table\,\ref{table:fluxes} (and, for 5.0 and 5.5 GHz,
plotted in Figure\,\ref{fig:Radio_curves}) but we could only place
(3$\sigma$) upper limits of $\lesssim 6$\,mJy ($\lesssim 30-50$\,\%)
on polarization in the 1.5\,GHz band, for the 3 epochs of 1.5\,GHz
observations.

Faraday rotation in the local or interstellar medium causes a rotation
of the polarization vectors at wavelength, $\lambda$, such that the
intrinsic electric vector position angle (EVPA) of the source is
related to the observed polarization angle, {\it PA}, by ${\rm EVPA} =
PA - RM\lambda^2$. The rotation measure is given by $RM \propto
\int_0^d \!n_{{\rm e}} B_{||} \,{\rm d}l$, where $n_{{\rm e}}$ is the
electron number density, $B_{||}$ is the magnetic field strength
parallel to the line of sight and $d$ is the distance to the source
(e.g.,
\citealt{Saikia-Salter1988:ARA&A.26,Johnston-Hollitt2004mim,Schnitzeler2010:MNRAS.409}).
Given observed polarization angles at multiple wavelengths, the
rotation measures and EVPAs (plotted in
Figure\,\ref{fig:Radio_curves}) are derived from a linear fit of $PA$
versus $\lambda^2$.


\section{Results and discussion}\label{section:discussion}

\subsection{Light curves, spectral indices \& polarization}\label{section:discribe}

As the radio light curves at frequencies $\geq 5$\,GHz
(Figure\,\ref{fig:Radio_curves}) share a similar morphology over the
sampled epochs, we consider only the most well-sampled radio frequency
(5 and 5.5\,GHz) for comparison to the X-ray light curves
(Figure\,\ref{fig:broad_lc}). The radio light curves, even though
under-sampled compared to the X-ray data, display a higher level of
large-scale variability (we cannot comment here on the smaller-scale,
shorter-time-scale variability due to the limited sampling of the
radio light curves).
The broadband light curves peak initially in the hard X-rays
(15--50\,keV) at MJD$\sim56188$, followed by the radio at
MJD$\sim56195$ and only then by the soft X-rays (0.3--10\,keV) at
MJD$\sim56200\mbox{--}56210$. During the rise of the soft X-ray emission, the
X-ray source softens to a hard intermediate state
\citep{Belloni2012:ATel.4450} and the soft X-ray hardness ratio
(1.5--10 keV/0.3--1.5 keV) reaches its minimum value
(Figure\,\ref{fig:broad_lc}, third panel).
After the initial rise to $\sim 30$mJy at MJD$\sim56195$, the radio
light curve displays a weaker peak of $\sim 10$mJy at MJD$\sim56214$
and a bright ``flare'' of $\sim 30$mJy at MJD$\sim56233$. We note,
given the rise time of the flare ($\lesssim 1.8$\,days) and the light
curve sampling, that other similar flares could have occurred but gone
undetected; demonstrating the importance of high-cadence radio
observations of outbursts.
In contrast to the radio light curve, neither the hard, nor soft,
X-ray bands exhibit any corresponding variation in flux.
When the X-ray light curve -- at least at hard energies -- does
display variability from approximately MJD$\sim56245$ until
MJD$\sim56270$ there is little or no radio data with which to compare.
At late times (around MJD 56300), after observations of the source had
become unconstrained by the Sun position, the fluxes of both radio and
hard X-rays were observed to have increased from the previous
measurements. By MJD 56325 the X-ray hardness had increased towards
its initial values, indicating that the source had reverted to the
canonical hard state from the hard intermediate state.


The radio light curve at 1.8\,GHz clearly exhibits a different
morphology from the higher radio frequencies.  Though its exact form
is difficult to constrain because of the sampling and uncertainties,
it seems to plateau, rather than peak, at MJD$\sim56195$ or possibly
later. This is similar behaviour to that exhibited by the soft X-rays
(0.3--10\,keV) but because of the sampling a more detailed comparison
is not possible. It is not uncommon to observe such later peaking and
flatter light curves at lower frequencies in synchrotron flaring
sources (e.g.,
\citealt{Miller-Jones2009:MNRAS.394,Armstrong2013:MNRAS.433}). The
lower frequency emission is expected to peak at greater radii (later
times) in an expanding jet (e.g.,
\citealt{vanderLaan1966:Natur.211,Hjellming1988:ApJ.328}) causing the
light curve to be smoothed out, and producing a low-frequency lag.

\begin{figure} 
  \centering 
   \resizebox{\hsize}{!}{\includegraphics[angle=-90]{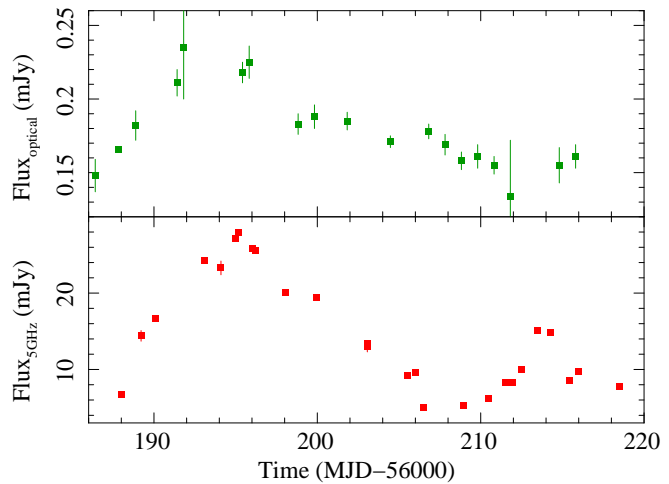} }
\vspace{10mm} 
 \caption{Optical ($i^{\prime}$-band, uncorrected for reddening) light curve
    \citep{Munoz-Darias2013:MNRAS.432} and the 5, 5.5\,GHz light curves
    over the same period for comparison.  }
 \label{fig:optical_lc} 
\end{figure}


Due to the relatively sparsely sampled optical
\citep{Munoz-Darias2013:MNRAS.432} and radio light curves, nothing more
than a cursory comparison is possible
(Figure\,\ref{fig:optical_lc}). While both peak at approximately the
same time, the radio increases more rapidly and by a larger factor
than the optical; likewise the radio decreases more quickly and by a
larger factor after that initial peak. The optical light curve
exhibits no discernible rise corresponding to the radio peak at
MJD$\sim56214$, nor to the preceding dip 6 days earlier (throughout
which, the radio spectrum is approximately flat).
This lack of a direct relationship between the two wavelengths implies
that if the jet makes a contribution to the optical flux it is not the
dominant emission mechanism (as has been observed in the
near-infrared, e.g.,
\citealt{Jain2001:ApJ554,Corbel2002:ApJ.573,Chaty2003:MNRAS.346,Russell2006:MNRAS.371,Russell2013:ApJ.768});
this is consistent with the detection by \citeauthor{Munoz-Darias2013:MNRAS.432}
of a strong H$\alpha$ emission line that they ascribe to
X-ray reprocessing in the accretion disk. However, we cannot rule out
that the jet might contribute to the optical flux and hence we cannot
place any limits on the frequency at which the jet transitions from
optically-thin to optically-thick emission.

As mentioned above, the multi-frequency radio light curves
(Figure\,\ref{fig:Radio_curves}) display similar morphologies as that
discussed for the 5 and 5.5\,GHz light curve. While the light curves
display high levels of variability, the spectral index over most of
the outburst is relatively stable at $\alpha \sim 0$, only deviating
from that flat spectrum at MJD$\gtrsim 56225$ and returning to $\alpha
\sim 0$ by MJD$\sim56300$. This decrease of $\alpha$ indicates a
steepening of the spectrum and occurs over the same period that the
level of fractional polarization increases to $\gtrsim 7\%$ at the two
frequencies where polarimetry is available. It should be noted that
the level of fractional polarization increases gradually, and
relatively steadily, over the entire period of the outburst sampled
from a starting level of $\sim 1\%$ to its maximum level of $\sim
50\%$.

The derived rotation measure of the interstellar medium seems to
display a high level of variation during the first 14 days; however,
since it is derived from only two frequencies, it may be inaccurate
and its variability exaggerated. While we are unable to confirm
variability we note that the observed polarization angles at each
frequency exhibit relative changes (Figure\,\ref{fig:Radio_curves},
lower panel) over the same period that the rotation measure changes
most significantly, suggesting that the variation is real. At epochs
between MJD 56199 and 56244 the rotation measure is consistent with a
value of $347 \pm 17$\,rad\,m$^{-2}$ which is comparable to the values
found by \cite{Roy2008:A&A.478}
for extragalactic sources in this region ($RM \approx
200\mbox{--}1000$ rad\,m$^{-2}$); at other epochs, assuming the
variability is real, the rotation measure is significantly lower. This
suggests a change of the magnetic field strength parallel to the line
of sight ($B_{||}$) or a change of the electron density ($n_{{\rm
    e}}$) in the region of the source. The EVPA displays a steady
shift from the initial angle of $\sim -80$\arcdeg, peaking at the same
time as the peak in flux, before becoming relatively stable at
approximately 0\arcdeg\ to 30\arcdeg. At late times ($>$ MJD 56240) it
displays another $\sim 90$\arcdeg\ rotation back to its approximate
initial position.


\subsubsection{Interpretation}\label{sect:interpretation}

The flat radio spectrum of $\alpha \sim0$ and the low level of
polarization ($\lesssim 10\%$) up to MJD 56225 -- spanning both the
rise and second peak -- are consistent with the radio emission
originating from a compact, self-absorbed jet which is expected to
exist in the hard state. Assuming a uniform magnetic field, such a jet
is expected to have a maximum fractional polarization, $FP \lesssim
300/(6p+13) \approx 11\%$ \citep{Longair1994:hea2.book} using the
canonical value of the electron energy distribution index, $p=2.4$.
This implies that, even though the X-ray hardness ratio had dropped
somewhat, the source did not make a full transition to a soft, thermal
dominant, state where the compact jets are expected to be quenched;
this is consistent with what was found from the X-ray observations
\citep{Belloni2012:ATel.4450}. In the hard state of other \lmxbs\ the
compact, self-absorbed jets are observed to be relatively steady and
correlated to the X-ray emission
(e.g. \citealt{Corbel2013:MNRAS.428}), though it often becomes
unstable as the source softens \citep{Fender2004:MNRAS.355}. Here we
see variability by up to a factor of $\sim10$, uncorrelated to the
X-ray emission, in an intermediate state which we will discuss further
in section \ref{sect:correlation}.

The radio data from MJD 56225 to 56250, comprising the ``flare'' at
MJD$\sim56233$ display greatly steepened radio spectral indices
($\ll0$) and a high level of polarization, which rises to values
inconsistent with those expected from self-absorbed emission but
consistent with those of optically-thin emission (where $FP \lesssim
100(p+1)/(p+7/3) \approx 72\%$; \citealt{Longair1994:hea2.book}).
This implies that the flare is an optically-thin ejection event in
contrast to optically-thick/self-absorbed peaks at earlier times.
It has recently been shown that the radio spectral index may display
steep values due to a quenching of higher radio frequencies near the
hard to soft state transition
\citep{Corbel2013:MNRAS.431,Russell2013:ApJ.768,vanderHorst2013:arXiv1308}.
However, the spectrum of this flare exhibits no such deviation from a
power law so there is no evidence that the spectral steepening is not
due to optically-thin ejecta. Additionally, such quenching would not
cause the observed increase in flux over the flare. 
While the discrete ejection is expected in the intermediate state, as
the source crosses the \lq jet line'
\citep{Fender2004:MNRAS.355,Fender2009:MNRAS.396}, it is interesting
to note that this is the first time that such an event has been
observed in a failed outburst.
It should be cautioned that, since the X-ray data do not span the
entire period when the jet is optically thin, we cannot rule out that
the source may have made a transition to a full soft state. Further
analysis of X-ray data from \swift\ and \integral\ is necessary to
confirm this but seems unlikely to support such a transition (Del Santo
et al., in preparation).

\begin{figure*}
  \centering 
  \resizebox{\hsize}{!}{\includegraphics[angle=0]{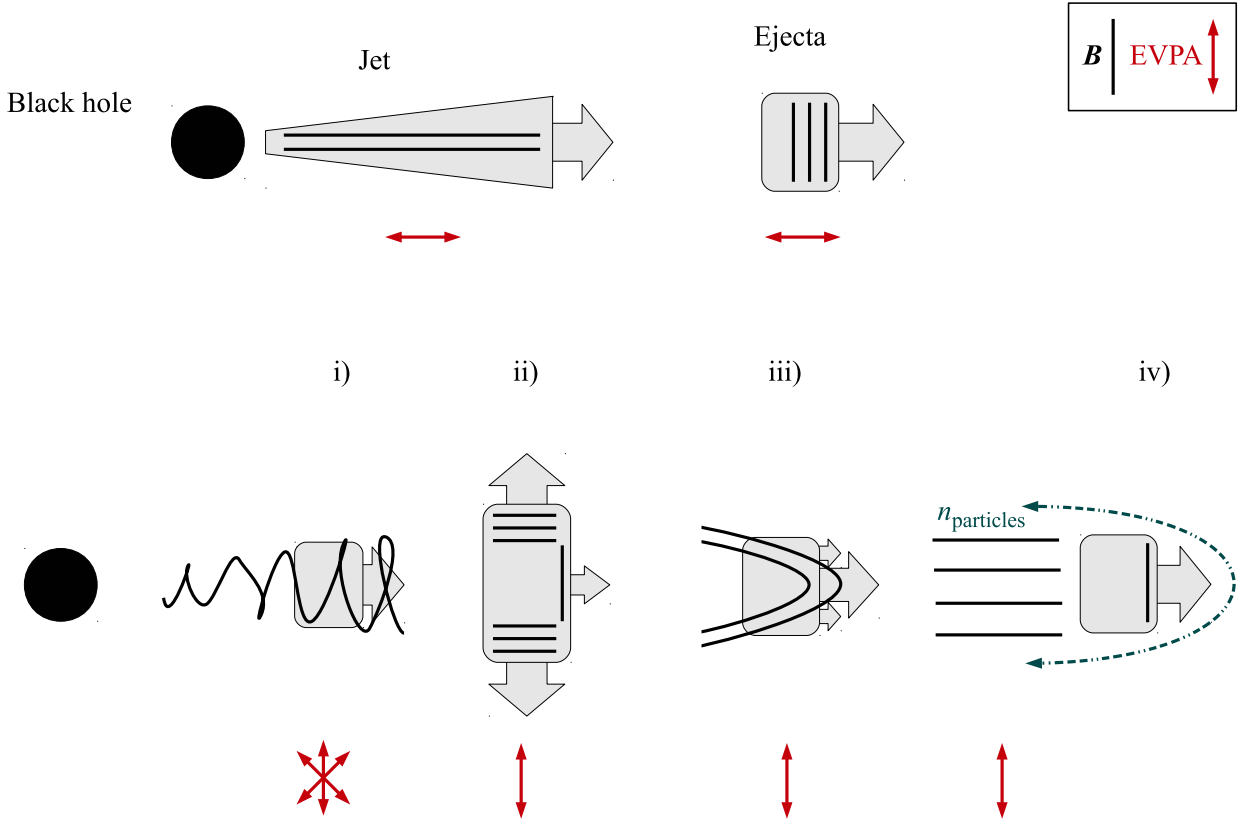} }
  \caption{Schematic of the various magnetic field (\textbf{\emph{B}})
    geometries, electric vector position angles (EVPA) and bulk
    motions (represented by arrow orientation and size) discussed in
    section \ref{sect:EVPA}. Top: the basic magnetic field geometry
    along the jet and at the ejecta shock front due to
    compression. Bottom: four possible geometries due to i) a helical
    field, ii) lateral expansion and compression, iii) velocity shear
    and iv) bow shock. All these may cause a dominant magnetic field
    orientation, and hence EVPA, that may deviate significantly from
    that expected in the simplest geometry.
}
  \label{fig:cartoon} 
\end{figure*}

\subsection{Electric vector position angle}\label{sect:EVPA}

The intrinsic EVPA displays a rotation of $\sim90$\arcdeg\ at early
times while the jet is self-absorbed (low level of polarization,
$\lesssim 10\%$, and a flat spectrum) and returns to initial values
when the jet displays evidence of changing to being optically thin
(steepening spectral slope and increased level of polarization).  The
latter rotation coincides with a significant reduction of the rotation
measure while the earlier rotation occurs during a period of variable
RM due to, e.g., a variation of the magnetic field strength ($B_{||}$)
or electron density ($n_{{\rm e}}$) in the region of the source.

Due to absorption effects the EVPA of the steady, self-absorbed jet is
expected to be aligned parallel to the magnetic field, which is in
turn, expected to be parallel to the jet axis (e.g.,
\citealt{Ginzburg1969:ARA&A.7,Longair1994:hea2.book}). Hence the EVPA
of $\sim0\mbox{--}30$\arcdeg\ from MJD 56195 to 56240 is a good
indicator that the jet is approximately orientated in the
North-northeast direction. This, however, is impossible to confirm
without high spatial resolution imaging, which is made difficult towards
the Galactic Centre by angular broadening at low radio
frequencies. 
The variable EVPA at the onset of the outburst may not, and likely
does not, indicate changes of the jet orientation, which we assume is
relatively stable at the core. Variable EVPAs, or \lq rotator events',
have previously been observed in the \lmxbs\ GRO\,J1655$-$40
\citep{Hannikainen2000:ApJ.540} and GRS\,1915+105
\citep{Fender2002:MNRAS.336}, and in a number of AGN (see
\citealt{Saikia-Salter1988:ARA&A.26}, and references therein) but are
not thought to indicate a physical rotation of the jet. Instead they
are thought to be caused by changes in the magnetic field or shock
conditions -- which, as demonstrated by the variability of the RM, is
clearly applicable here -- or possibly by a twisted or helical
magnetic field (e.g.,
\citealt{Gomez2008:ApJ.681,Marscher2008:Natur.452}).

The EVPA of the optically-thin ejecta is expected to be aligned
perpendicular to the magnetic field (e.g.,
\citealt{Ginzburg1969:ARA&A.7,Longair1994:hea2.book}). In the simplest
geometry, where the dominant magnetic field is due to shock
compression \citep{Laing1980:MNRAS.193}, the field is parallel to the
shock front (i.e., perpendicular to the jet axis) and hence the EVPA
is parallel to the jet axis, as it was in the self-absorbed
case. Thus, the rotation we observe when the jet becomes optically
thin is not expected in this basic description.
However, the emission from the discrete ejecta may not be described by
this simple geometry but by local or large scale variations of the
magnetic field or particle density near the source that may have much
more complex structures (Figure\,\ref{fig:cartoon}).  For example,
large scale helical magnetic fields may dominate at large distances
from the black hole (e.g.,
\citealt{Gomez2008:ApJ.681,Marscher2008:Natur.452}); lateral expansion
of the ejecta may produce (via compression) a dominant magnetic field
parallel to the bulk motion of the ejecta; the field may be elongated
due to velocity shear \citep{Aloy2000:ApJ.528}; knots or hotspots
might exist, where the local magnetic field and particle density is
compressed or distorted by a bow shock and particles flow along the
shock front to illuminate the jet-aligned magnetic field behind the
shock \citep{Dreher1987:ApJ.316}.
These field variations can also explain increases (or decreases,
depending on magnetic field direction) in the measured RM at late
times. Such spatially dependent magnetic variations have been directly
imaged and resolved by interferometry in both \lmxbs\ (e.g.,
\citealt{Miller-Jones2008:ApJ.682}) and, more commonly, AGN (e.g.,
\citealt{Lister2005:AJ.130,Gomez2008:ApJ.681,Homan2009:ApJ.696}), but
it is not possible to do so for the unresolved jets of
\swiftseventeen.


\begin{figure*}
  \centering 
  \resizebox{14cm}{!}{\includegraphics[angle=0]{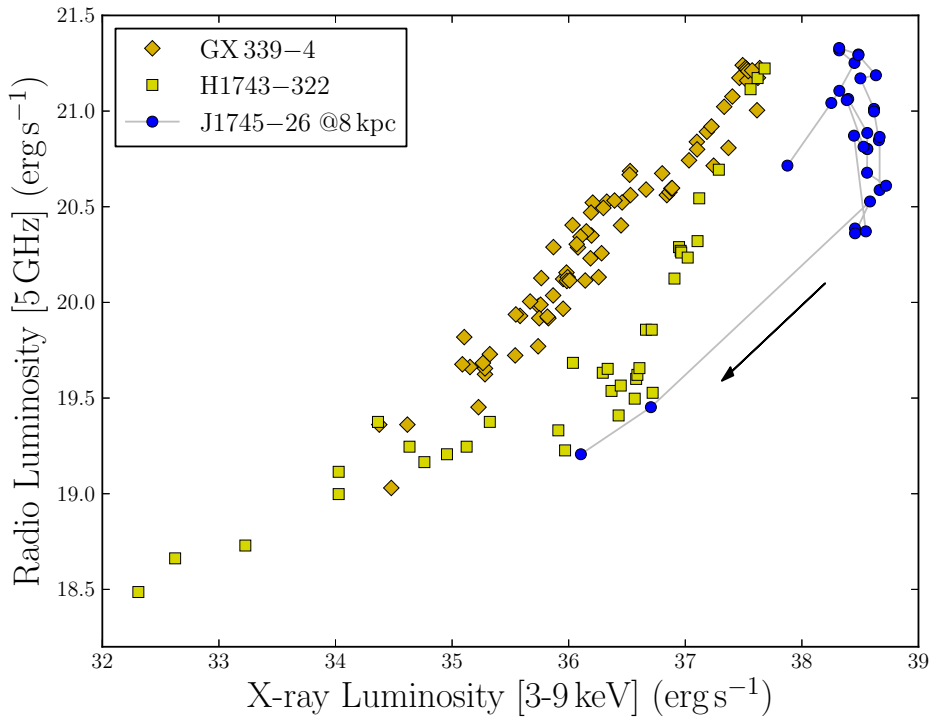}}
  \vspace{0mm} 
  \caption{The hard-state X-ray -- radio luminosity correlation shows
    representative data from the upper (GX\,339$-$4;
    \citealt{Corbel2013:MNRAS.428}) and lower (H1743$-$322;
    \citealt{Coriat2011:MNRAS.414}) branches, overlaid by our data for
    \swiftseventeen\ at an assumed distance of 8\,kpc (the arrow
    represents the direction of temporal evolution).
}
  \label{fig:Lx-Lr} 
\end{figure*}

\subsection{Flare energetics}\label{sect:energetics}

While there are 3 clear peaks in the 5 and 5.5\,GHz light curve at MJD
$\sim56195$, $\sim56214$ and $\sim56233$ (Figure\,\ref{fig:broad_lc}),
we identify only the last as a flare, or discrete ejection event,
because of its optically-thin spectrum and high fractional
polarization. Using the approximations and
formulation of \cite{Fender2006:csxs.book381}, and assuming
equipartition (i.e., the energy is approximately equally divided
between emitting electrons and the magnetic field), the minimum
internal energy required to launch a discrete flaring event is
\[
E_{\rm min} \sim 7\times10^{39} 
\left(\frac{\Delta t}{\rm d}\right)^{9/7} 
\left(\frac{\nu}{\rm GHz}\right)^{2/7}
\left(\frac{F_{\nu}}{\rm mJy}\right)^{4/7} 
\left(\frac{d}{\rm kpc}\right)^{8/7} {\rm erg},
\] 
where $\Delta t$ is the rise time and $d$ is distance to the source.
The related mean power  of the ejection event is 
$P_{\rm min} = E_{\rm min} / \Delta t$,
 the magnetic field strength at minimum energy is 
\[
B_{\rm eq} \sim 2
\left(\frac{\Delta t}{\rm d}\right)^{-6/7} 
\left(\frac{\nu}{\rm GHz}\right)^{1/7} 
\left(\frac{F_{\nu}}{\rm mJy}\right)^{2/7} 
\left(\frac{d}{\rm kpc}\right)^{4/7}  {\rm mG}, 
\]
and the corresponding Lorentz factor of the synchrotron emitting electrons is 
\[
\gamma_e \sim 950 \left(\frac{\nu}{\rm GHz}\right)^{1/2} \left(\frac{B}{\rm mG}\right)^{-1/2}. 
\]
Because of the approximately linear dependency ($E_{\rm min} \propto
d^{1.14}$), the unknown distance to this source (see section
\ref{sect:correlation}) will only have a modest effect when
calculating the minimum energy. For a rise time of 1.83 days and
distances in the range 5 to 8.5 kpc, we find $E_{\rm min} \sim
10^{42}$\,erg and $P_{\rm min} \sim 10^{37}$\,erg~s$^{-1}$, which
imply an equipartition magnetic field strength of $\sim10$\,mG and
emitting electrons of Lorentz factors, $\gamma_{{\rm e}} \sim650$.
Without constraints on the bulk motion (i.e., bulk Lorentz factor,
$\Gamma$) of the ejecta, we cannot correct for Doppler (de-)boosting
but the estimated power is comparable to those of other \lmxb\ sources
which range from $\sim 10^{36} \mbox{--} 10^{39}$\,erg~s$^{-1}$
\citep{Fender1999:MNRAS.304,Brocksopp2007:MNRAS.378,Brocksopp2013:MNRAS.432}.

Assuming a representative black hole mass of
8\Msun\ \citep{Kreidberg2012:ApJ.757}, the Eddington luminosity is
$L_{{\rm Edd}} \approx 1.3 \times 10^{38} (M/\Msun)$ erg~s$^{-1} = 1
\times 10^{39}$ erg~s$^{-1}$.  The observed soft X-ray (1--10\,keV)
flux at the time of the flare, $F_{{\rm X}}\approx 1 \times 10^{-7}$
erg~cm$^{-2}$~s$^{-1}$, corresponds to X-ray luminosities of $L_{{\rm
    X}} \approx 4\mbox{--}12 \times 10^{38}$ erg~s$^{-1}$ for
distances of 5--8.5 kpc. This implies that the source is at a
significant fraction of the Eddington luminosity (even without a
bolometric correction), though it does not seem to have transitioned
to the soft state at the time of the discrete ejection event. This is
consistent with the hard to soft state transition X-ray luminosities
inferred by \cite{Dunn2010:MNRAS.403} that approach $L_{{\rm Edd}}$
and are an order of magnitude greater than the inferred soft to hard
state transition luminosities. However, further discussion requires a
more detailed X-ray spectral analysis, beyond the scope of this work.


\subsection{Radio--X-ray correlation}\label{sect:correlation}

We plot the hard-state X-ray--radio luminosity correlation (see
section \ref{section:introduction}) using representative data from the
upper (GX\,339$-$4; \citealt{Corbel2013:MNRAS.428}, assuming, as
they do, a distance of 8\,kpc; \citealt{Zdziarski2004:MNRAS.351}) and
lower (H1743$-$322; \citealt{Coriat2011:MNRAS.414}) branches
(Figure\,\ref{fig:Lx-Lr}).
For comparison, we calculate the luminosities of
\swiftseventeen\ over/at the same energies as
\cite{Corbel2013:MNRAS.428} and \cite{Coriat2011:MNRAS.414}, at
various distances (though we only plot 8\,kpc for clarity) and only at
epochs where X-ray observations occurred within 24 hours of the radio
observations (the arrow represents the direction of temporal
evolution).

The distance to \swiftseventeen\ is highly uncertain; based on optical
and X-ray fluxes at the peak of outburst,
\cite{Munoz-Darias2013:MNRAS.432} suggest a, poorly constrained,
distance of 1--7\,kpc and given the source position, close to the
Galactic centre, it is unlikely to be further than 8.5\,kpc.
We find that at all these distances the luminosities of the source
fall below the correlation, except for the two late-time points (MJD
56328 and 56377) that fall on the lower branch for distances
7-8.5\,kpc. By comparison with Figure \ref{fig:broad_lc}, these two
points are the only points on the correlation that occur when the
observed hardness ratio displays a steady, high value (which may
indicate that the system is in the canonical hard state). All other
points occur as the X-rays are rising to peak and the hardness ratio
is softening (MJD 56187--56200), or during the intermediate state when
the X-rays are softest (MJD 56200--56232); the majority of the points
on the plot are likely deviant from the correlation because they do
not occur during the hard state. The lack of hard-state luminosities
makes it difficult to confirm whether the points fall on the
hard-state correlation and hence to use the correlation to constrain
the distance to \swiftseventeen.

The form of the deviation from the correlation is similar to that
observed for GX\,339$-$4, which was interpreted as the jet turning on
and off in the intermediate state \citep{Corbel2013:MNRAS.428}.  While
the jet of \swiftseventeen\ is never fully quenched, this deviation
further confirms that the source was in an intermediate, as opposed to
hard state, for the majority of our observations. In this state, the
compact jet (the flare is not sampled due to a lack of simultaneous
X-ray observations) displayed a high level of light curve variability
suggesting an unstable jet with variable power.  Unlike GX\,339$-$4,
which shows an X-ray hysteresis between the on/off deviation, we
observe no such difference (similar to MAXI J1659$-$152;
\citealt{vanderHorst2013:arXiv1308}); this is because GX\,339$-$4
evolved through the stages of the standard hardness intensity diagram
(e.g.,
\citealt{homan2001:ApJS132,homan_2005ApSS...300,belloni_2010LNP...794})
while, as a failed outburst, \swiftseventeen\ did not. Such an
interpretation is independent of the distance to the source and
consistent with our interpretation of the radio light curves and
spectral indices.


\section{Conclusion}\label{section:conclusion}

We obtained multi-frequency radio data of the black hole
\lmxb, \swiftseventeen\ from the VLA, ATCA and KAT-7 radio arrays
during the failed outburst of 2012--2013. For the majority of the
observations, the radio spectral and polarimetric data are consistent with
self-absorbed synchrotron emission from a compact jet and, as
expected, the X-ray light curves and hardness ratios imply a source in
the canonical hard or intermediate state.
The radio light curves display a high level of large scale variability
compared to the X-ray emission, and deviations from the standard
radio--X-ray correlations, which we interpret as the unstable compact
jet's energy decreasing and increasing in the intermediate state
(without ever fully quenching). From the measured Stokes polarimetry
parameters we infer the orientation of the unresolved jet as being
approximately North-northeast.

For a period the radio data display strong evidence -- steep spectral
index and a very high level of polarization -- for a discrete ejection
event. While discrete ejecta are expected in the intermediate state,
as the source crosses the \lq jet line', it is interesting to note
that this is the first time that such an event has been observed in a
failed outburst where the soft state has not been reached.
Though the distance to the source is poorly
constrained, for a range of physically realistic distances we find, as
expected, that the minimum energy required to launch such a discrete
flaring event is a significant proportion of the Eddington luminosity,
which suggests a very high level of accretion. Coincident with the
discrete ejection event, we observe a rotation of the electric vector
position angle inferring a complex and variable magnetic field
geometry in the vicinity of the source.
Future, multi-wavelength analysis of this outburst, particularly at
X-ray and optical wavelengths, that can constrain the physical
parameters of the accretion disk, should allow the state of the system
to be accurately determined and hence better describe the relationship
between the observed radio jet and the accretion disk.


\section*{Acknowledgements}

We thank A.H. Bridle for useful discussions on magnetic field
geometries and J.P. Macquart for useful discussions on the estimates
of polarization. We thank T. Hovatta for her assistance with OVRO
observations.
This work was supported by Australian Research Council grant
DP120102393.
MC and RPA acknowledge the financial assistance of the National
Research Foundation (NRF) through an SKA SA Fellowship.
GRS is supported by an NSERC Discovery Grant.
SC acknowledges the financial support from the UnivEarthS Labex
program of Sorbonne Paris Cit\'e (ANR-10-LABX-0023 and
ANR-11-IDEX-0005-02).
DMR acknowledges support from a Marie Curie Intra European Fellowship
within the 7th European Community Framework Programme under contract
no. IEF 274805.
The Australia Telescope Compact Array is part of the Australia
Telescope National Facility which is funded by the Commonwealth of
Australia for operation as a National Facility managed by CSIRO.
The National Radio Astronomy Observatory is a facility of the National
Science Foundation operated under cooperative agreement by Associated
Universities, Inc.
This research has made use of NASA's Astrophysics Data System and the
SIMBAD database, operated at CDS, Strasbourg, France.
Swift XRT data was supplied by the UK Swift Science Data Centre at the
University of Leicester and Swift BAT transient monitor results were
provided by the Swift/BAT team.


\label{lastpage}

\begin{table*}
  \centering \contcaption{Radio flux densities of source, $F_{\nu}$,
    at frequency, $\nu$, and Stokes $Q$ and $U$ flux densities at that
    frequency (all given before systematic errors are added).}
\begin{tabular}{l l l l l} 
  \hline
  Epoch   & $\nu$  & $F_{\nu}$     & $Q$     & $U$  \\
  (MJD)   & (GHz) & (mJy) & (mJy beam$^{-1}$) & (mJy beam$^{-1}$)  \\
  \hline
56187.991 & 5.0 & 6.77 $\pm$ 0.13   & -0.025 $\pm$ 0.021 & -0.113 $\pm$ 0.021 \\ 
56187.991 & 7.5 & 6.26 $\pm$ 0.03   & -0.075 $\pm$ 0.012 & -0.074 $\pm$ 0.012 \\ 
56189.210 & 5.5 & 14.40 $\pm$ 0.70   & ... & ...  \\ 
56189.210 & 9.0 & 13.70 $\pm$ 0.60   & ... & ...  \\ 
56190.087 & 5.0 & 16.66 $\pm$ 0.06   & 0.063 $\pm$ 0.022 & -0.206 $\pm$ 0.024 \\ 
56190.087 & 7.5 & 15.92 $\pm$ 0.07   & 0.035 $\pm$ 0.023 & -0.205 $\pm$ 0.025 \\ 
56191.327 & 1.90 & 19 $\pm$ 12   & ... & ...  \\ 
56193.071 & 5.0 & 24.24 $\pm$ 0.08   & -0.138 $\pm$ 0.050 & 0.253 $\pm$ 0.055 \\ 
56193.071 & 7.5 & 23.52 $\pm$ 0.11   & 0.182 $\pm$ 0.055 & 0.508 $\pm$ 0.053 \\ 
56193.071 & 20.8 & 26.78 $\pm$ 0.08   & ... & ...  \\ 
56193.071 & 25.9 & 26.73 $\pm$ 0.07   & ... & ...  \\ 
56194.100 & 5.5 & 23.30 $\pm$ 0.90   & ... & ...  \\ 
56194.100 & 9.0 & 23.10 $\pm$ 0.70   & ... & ...  \\ 
56195.016 & 5.0 & 27.11 $\pm$ 0.35   & 0.006 $\pm$ 0.060 & 0.581 $\pm$ 0.062 \\ 
56195.016 & 7.5 & 26.69 $\pm$ 0.39   & 0.101 $\pm$ 0.051 & 0.577 $\pm$ 0.052 \\ 
56195.016 & 20.8 & 28.82 $\pm$ 0.21   & ... & ...  \\ 
56195.016 & 25.9 & 28.74 $\pm$ 0.15  & ... & ...  \\ 
56195.160 & 5.5 & 27.90 $\pm$ 0.20   & ... & ...  \\ 
56195.160 & 9.0 & 28.10 $\pm$ 0.30   & ... & ...  \\ 
56195.336 & 1.90 & 24.51 $\pm$ 8.53   & ... & ...  \\ 
56196.023 & 1.35 & 22.54 $\pm$ 0.30   & ... & ...  \\ 
56196.023 & 1.41 & 21.74 $\pm$ 0.33   & ... & ...  \\ 
56196.023 & 1.47 & 20.17 $\pm$ 0.38   & ... & ...  \\ 
56196.023 & 1.72 & 22.21 $\pm$ 0.39   & ... & ...  \\ 
56196.023 & 1.78 & 21.03 $\pm$ 0.36   & ... & ...  \\ 
56196.023 & 1.85 & 20.95 $\pm$ 0.41   & ... & ...  \\ 
56196.023 & 1.91 & 23.38 $\pm$ 0.48   & ... & ...  \\ 
56196.023 & 1.97 & 23.48 $\pm$ 0.81   & ... & ...  \\ 
56196.023 & 5.0 & 25.82 $\pm$ 0.10   & -0.001 $\pm$ 0.055 & 0.513 $\pm$ 0.053 \\ 
56196.023 & 7.5 & 25.26 $\pm$ 0.16   & -0.046 $\pm$ 0.052 & 0.768 $\pm$ 0.054 \\ 
56196.023 & 20.8 & 28.35 $\pm$ 0.26   & ... & ...  \\ 
56196.023 & 25.9 & 27.90 $\pm$ 0.27   & ... & ...  \\ 
56196.023 & 31.5 & 31.32 $\pm$ 0.09   & ... & ...  \\ 
56196.023 & 37.5 & 36.46 $\pm$ 0.58   & ... & ...  \\ 
56196.023 & 41.5 & 38.9 $\pm$ 1.4   & ... & ...  \\ 
56196.023 & 47.5 & 41.0 $\pm$ 1.5   & ... & ...  \\ 
56196.200 & 5.5 & 25.60 $\pm$ 0.40   & ... & ...  \\ 
56196.200 & 9.0 & 25.90 $\pm$ 0.40   & ... & ...  \\ 
56196.335 & 1.90 & 24 $\pm$ 9   & ... & ...  \\ 
56197.334 & 1.90 & 25 $\pm$ 8   & ... & ...  \\ 
56198.039 & 1.35 & 20.02 $\pm$ 0.28   & ... & ...  \\ 
56198.039 & 1.41 & 22.16 $\pm$ 0.33   & ... & ...  \\ 
56198.039 & 1.47 & 20.39 $\pm$ 0.50   & ... & ...  \\ 
56198.039 & 1.72 & 21.19 $\pm$ 0.72   & ... & ...  \\ 
56198.039 & 1.78 & 22.91 $\pm$ 0.48   & ... & ...  \\ 
56198.039 & 1.85 & 21.42 $\pm$ 0.66   & ... & ...  \\ 
56198.039 & 1.91 & 20.91 $\pm$ 0.63   & ... & ...  \\ 
56198.039 & 1.97 & 22 $\pm$ 2   & ... & ...  \\ 
56198.039 & 5.0 & 20.10 $\pm$ 0.09   & -0.217 $\pm$ 0.057 & 0.405 $\pm$ 0.058 \\ 
56198.039 & 7.5 & 18.70 $\pm$ 0.07   & 0.115 $\pm$ 0.053 & 0.592 $\pm$ 0.052 \\ 
56198.039 & 14.4 & 19.39 $\pm$ 0.07   & 0.467 $\pm$ 0.067 & 0.430 $\pm$ 0.068 \\ 
56198.039 & 17.2 & 19.67 $\pm$ 0.08   & 0.594 $\pm$ 0.097 & 0.597 $\pm$ 0.099 \\ 
56198.039 & 20.8 & 19.62 $\pm$ 0.37   & ... & ...  \\ 
56198.039 & 25.9 & 20.25 $\pm$ 0.46   & ... & ...  \\ 
56198.039 & 41.5 & 18.82 $\pm$ 0.20   & ... & ...  \\ 
56198.039 & 47.5 & 21.1 $\pm$ 1.1   & ... & ...  \\ 
56198.349 & 1.90 & 26 $\pm$ 6   & ... & ...  \\ 
56199.929 & 1.35 & 20.48 $\pm$ 0.61   & ... & ...  \\ 
56199.929 & 1.41 & 21.15 $\pm$ 0.89   & ... & ...  \\ 
\hline
\end{tabular}
\end{table*}

\begin{table*}  
  \centering            
  \contcaption{}                        
\begin{tabular}{l l l l l} 
  \hline
  Epoch   & $\nu$  & $F_{\nu}$     & $Q$     & $U$  \\  
  (MJD)   & (GHz) & (mJy) & (mJy beam$^{-1}$) & (mJy beam$^{-1}$)  \\
  \hline  
56199.929 & 1.47 & 20.09 $\pm$ 0.59   & ... & ...  \\ 
56199.929 & 1.72 & 20.7 $\pm$ 1.8   & ... & ...  \\ 
56199.929 & 1.78 & 16.91 $\pm$ 0.48   & ... & ...  \\ 
56199.929 & 1.85 & 17.29 $\pm$ 0.53   & ... & ...  \\ 
56199.929 & 1.91 & 19.42 $\pm$ 0.77   & ... & ...  \\ 
56199.929 & 5.0 & 19.36 $\pm$ 0.05   & -0.448 $\pm$ 0.051 & 0.094 $\pm$ 0.051 \\ 
56199.929 & 7.5 & 18.40 $\pm$ 0.04   & 0.303 $\pm$ 0.049 & 0.363 $\pm$ 0.049 \\ 
56199.929 & 14.4 & 18.28 $\pm$ 0.12   & 0.300 $\pm$ 0.103 & 0.105 $\pm$ 0.101 \\ 
56199.929 & 17.2 & 18.78 $\pm$ 0.18   & 0.291 $\pm$ 0.118 & 0.473 $\pm$ 0.115 \\ 
56199.929 & 14.4 & 17.32 $\pm$ 0.11   & ... & ...  \\ 
56199.929 & 17.2 & 16.64 $\pm$ 0.25   & ... & ...  \\ 
56199.929 & 20.8 & 17.32 $\pm$ 0.11   & ... & ...  \\ 
56199.929 & 25.9 & 14.49 $\pm$ 0.23   & ... & ...  \\ 
56199.929 & 41.5 & 0.00 $\pm$ 0.00   & ... & ...  \\ 
56199.929 & 47.5 & 0.00 $\pm$ 0.00   & ... & ...  \\ 
56200.383 & 1.90 & 29 $\pm$ 2   & ... & ...  \\ 
56201.335 & 1.90 & 28 $\pm$ 8   & ... & ...  \\ 
56202.348 & 1.90 & 27 $\pm$ 4   & ... & ...  \\ 
56203.044 & 5.0 & 13.41 $\pm$ 0.06   & -0.346 $\pm$ 0.054 & 0.204 $\pm$ 0.050 \\ 
56203.044 & 7.5 & 12.65 $\pm$ 0.03   & 0.010 $\pm$ 0.042 & 0.392 $\pm$ 0.046 \\ 
56203.044 & 20.8 & 13.27 $\pm$ 0.13   & ... & ...  \\ 
56203.044 & 25.9 & 12.87 $\pm$ 0.24   & ... & ...  \\ 
56203.060 & 5.5 & 13.00 $\pm$ 0.70   & ... & ...  \\ 
56203.060 & 9.0 & 11.80 $\pm$ 0.40   & ... & ...  \\ 
56203.369 & 1.90 & 15 $\pm$ 9   & ... & ...  \\ 
56205.490 & 5.5 & 9.20 $\pm$ 0.10   & ... & ...  \\ 
56205.490 & 9.0 & 7.60 $\pm$ 0.10   & ... & ...  \\ 
56206.039 & 5.0 & 9.57 $\pm$ 0.05   & -0.349 $\pm$ 0.051 & 0.021 $\pm$ 0.048 \\ 
56206.039 & 7.5 & 8.76 $\pm$ 0.04   & -0.109 $\pm$ 0.044 & 0.369 $\pm$ 0.045 \\ 
56206.039 & 20.8 & 8.66 $\pm$ 0.06   & ... & ...  \\ 
56206.039 & 25.9 & 8.46 $\pm$ 0.06   & ... & ...  \\ 
56206.490 & 5.5 & 5.05 $\pm$ 0.17   & ... & ...  \\ 
56206.490 & 9.0 & 3.60 $\pm$ 0.20   & ... & ...  \\ 
56208.955 & 5.0 & 5.32 $\pm$ 0.05   & -0.304 $\pm$ 0.060 & 0.093 $\pm$ 0.057 \\ 
56208.955 & 7.5 & 4.49 $\pm$ 0.05   & 0.096 $\pm$ 0.051 & 0.267 $\pm$ 0.048 \\ 
56208.955 & 20.8 & 4.42 $\pm$ 0.06   & ... & ...  \\ 
56208.955 & 25.9 & 4.33 $\pm$ 0.05   & ... & ...  \\ 
56210.444 & 1.90 & 13 $\pm$ 12   & ... & ...  \\ 
56210.480 & 5.5 & 6.22 $\pm$ 0.08   & ... & ...  \\ 
56210.480 & 9.0 & 4.30 $\pm$ 0.10   & ... & ...  \\ 
56211.490 & 5.5 & 8.28 $\pm$ 0.08   & ... & ...  \\ 
56211.490 & 9.0 & 7.53 $\pm$ 0.07   & ... & ...  \\ 
56211.965 & 20.8 & 7.04 $\pm$ 0.12   & ... & ...  \\ 
56211.965 & 25.9 & 7.11 $\pm$ 0.12   & ... & ...  \\ 
56211.967 & 5.0 & 8.27 $\pm$ 0.04   & -0.239 $\pm$ 0.046 & 0.143 $\pm$ 0.045 \\ 
56211.967 & 7.5 & 7.81 $\pm$ 0.04   & 0.073 $\pm$ 0.044 & 0.386 $\pm$ 0.044 \\ 
56212.380 & 1.90 & 8 $\pm$ 3   & ... & ...  \\ 
56212.490 & 5.5 & 10.04 $\pm$ 0.11   & ... & ...  \\ 
56212.490 & 9.0 & 9.15 $\pm$ 0.15   & ... & ...  \\ 
56213.470 & 5.5 & 15.11 $\pm$ 0.08   & ... & ...  \\ 
56213.470 & 9.0 & 12.40 $\pm$ 0.10   & ... & ...  \\ 
56214.260 & 5.5 & 14.90 $\pm$ 0.20   & ... & ...  \\ 
56214.260 & 9.0 & 13.74 $\pm$ 0.06   & ... & ...  \\ 
56214.417 & 1.90 & 16 $\pm$ 8   & ... & ...  \\ 
56215.440 & 5.5 & 8.50 $\pm$ 0.10   & ... & ...  \\ 
56215.440 & 9.0 & 9.90 $\pm$ 0.20   & ... & ...  \\ 
56215.962 & 20.8 & 9.99 $\pm$ 0.05   & ... & ...  \\ 
56215.962 & 25.9 & 9.98 $\pm$ 0.14   & ... & ...  \\ 
56215.963 & 5.0 & 9.71 $\pm$ 0.03   & -0.295 $\pm$ 0.045 & 0.303 $\pm$ 0.044 \\ 
56215.963 & 7.5 & 9.53 $\pm$ 0.04   & 0.097 $\pm$ 0.040 & 0.360 $\pm$ 0.040 \\ 
\hline
\end{tabular}
\end{table*}

\begin{table*}
  \centering	
  \contcaption{ } 
\begin{tabular}{l l l l l} 
  \hline
  Epoch   & $\nu$  & $F_{\nu}$     & $Q$     & $U$  \\ 
  (MJD)   & (GHz) & (mJy) & (mJy beam$^{-1}$) & (mJy beam$^{-1}$)  \\
  \hline
56218.490 & 5.5 & 7.75 $\pm$ 0.04   & ... & ...  \\ 
56218.490 & 9.0 & 6.83 $\pm$ 0.03   & ... & ...  \\ 
56220.370 & 5.5 & 7.70 $\pm$ 0.03   & ... & ...  \\ 
56220.370 & 9.0 & 6.34 $\pm$ 0.03   & ... & ...  \\ 
56223.360 & 5.5 & 3.07 $\pm$ 0.02   & ... & ...  \\ 
56223.360 & 9.0 & 2.24 $\pm$ 0.03   & ... & ...  \\ 
56224.996 & 5.0 & 3.18 $\pm$ 0.05   & -0.193 $\pm$ 0.056 & 0.125 $\pm$ 0.056 \\ 
56224.996 & 7.5 & 2.97 $\pm$ 0.03   & 0.107 $\pm$ 0.042 & 0.219 $\pm$ 0.041 \\ 
56224.996 & 20.8 & 3.10 $\pm$ 0.06   & ... & ...  \\ 
56224.996 & 25.9 & 2.99 $\pm$ 0.13   & ... & ...  \\ 
56225.150 & 5.5 & 3.00 $\pm$ 0.10   & ... & ...  \\ 
56225.150 & 9.0 & 2.20 $\pm$ 0.10   & ... & ...  \\ 
56231.160 & 5.5 & 4.40 $\pm$ 0.30   & ... & ...  \\ 
56231.160 & 9.0 & 3.50 $\pm$ 0.40   & ... & ...  \\ 
56232.995 & 5.0 & 26.33 $\pm$ 0.06   & -0.909 $\pm$ 0.058 & -0.947 $\pm$ 0.058 \\ 
56232.995 & 7.5 & 22.35 $\pm$ 0.05   & -0.942 $\pm$ 0.049 & 0.622 $\pm$ 0.049 \\ 
56232.995 & 20.8 & 14.67 $\pm$ 0.09   & ... & ...  \\ 
56232.995 & 25.9 & 13.17 $\pm$ 0.07   & ... & ...  \\ 
56235.330 & 5.5 & 7.18 $\pm$ 0.09   & ... & ...  \\ 
56235.330 & 9.0 & 4.63 $\pm$ 0.07   & ... & ...  \\ 
56242.204 & 1.90 & 21 $\pm$ 12   & ... & ...  \\ 
56243.210 & 5.5 & 3.60 $\pm$ 0.20   & ... & ...  \\ 
56243.210 & 9.0 & 2.00 $\pm$ 0.40   & ... & ...  \\ 
56243.827 & 5.0 & 3.37 $\pm$ 0.05   & 0.223 $\pm$ 0.053 & -0.121 $\pm$ 0.054 \\ 
56243.827 & 7.5 & 2.46 $\pm$ 0.05   & -0.024 $\pm$ 0.046 & -0.261 $\pm$ 0.046 \\ 
56243.827 & 20.8 & 1.31 $\pm$ 0.04   & ... & ...  \\ 
56243.827 & 25.9 & 0.67 $\pm$ 0.05   & ... & ...  \\ 
56248.782 & 5.0 & 0.50 $\pm$ 0.04   & -0.101 $\pm$ 0.062 & 0.225 $\pm$ 0.060 \\ 
56248.782 & 7.5 & 0.37 $\pm$ 0.02   & -0.126 $\pm$ 0.050 & 0.115 $\pm$ 0.050 \\ 
56248.782 & 20.8 & 2.64 $\pm$ 0.00   & ... & ...  \\ 
56248.782 & 25.9 & 2.55 $\pm$ 0.00   & ... & ...  \\ 
56303.080 & 5.5 & 0.85 $\pm$ 0.03   & ... & ...  \\ 
56303.080 & 9.0 & 0.89 $\pm$ 0.03   & ... & ...  \\ 
56327.790 & 5.5 & 0.37 $\pm$ 0.02   & ... & ...  \\ 
56327.790 & 9.0 & 0.38 $\pm$ 0.02   & ... & ...  \\ 
56344.830 & 5.5 & 0.23 $\pm$ 0.02   & ... & ...  \\ 
56344.830 & 9.0 & 0.21 $\pm$ 0.02   & ... & ...  \\ 
56377.850 & 5.5 & 0.21 $\pm$ 0.05   & ... & ...  \\ 
56377.850 & 9.0 & 0.19 $\pm$ 0.03   & ... & ...  \\ 
\hline
\end{tabular}
\end{table*}

\end{document}